\documentstyle[12pt]{article}
\newlength{\dinwidth}
\newlength{\dinmargin}
\newlength{\extraspace}
\newlength{\extraspaces}
\newcommand{\be}{\begin{equation}
\addtolength{\abovedisplayskip}{\extraspaces}
\addtolength{\belowdisplayskip}{\extraspaces}
\addtolength{\abovedisplayshortskip}{\extraspace}
\addtolength{\belowdisplayshortskip}{\extraspace}}
\newcommand{\ee}{\end{equation}}
\newcommand{\bdm}{\begin{displaymath}
\addtolength{\abovedisplayskip}{\extraspaces}
\addtolength{\belowdisplayskip}{\extraspaces}
\addtolength{\abovedisplayshortskip}{\extraspace}
\addtolength{\belowdisplayshortskip}{\extraspace}}
\newcommand{\edm}{\end{displaymath}}
\renewcommand{\thefootnote}{\fnsymbol{footnote}}
\def\simlt{\mathrel{\lower2.5pt\vbox{\lineskip=0pt\baselineskip=0pt
           \hbox{$<$}\hbox{$\sim$}}}}
\newcommand{\beq}{\begin{equation}}
\newcommand{\eeq}{\end{equation}}
\newcommand{\bea}{\begin{eqnarray}}
\newcommand{\eea}{\end{eqnarray}}
\newcommand{\ts}{\thinspace}
\newcommand{\semi}{; \\}
\newcommand{\pr}{Phys.\ Rev.\ }
\newcommand{\prp}{Phys.\ Rep.\ }
\newcommand{\prl}{Phys.\ Rev.\ Lett.\ }
\newcommand{\np}{Nucl.\ Phys.\ {\bf B}}
\newcommand{\pl}{Phys.\ Lett.\ {\bf B}}
\newcommand{\rmp}{Reviews of Modern Physics\ }
\newcommand{\peetee}{p_T}
\newcommand{\etoc}{\eta_8}
\newcommand{\etas}{\eta_0}
\newcommand{\rhoc}{\rho_8}
\newcommand{\ttbar}{t{\overline t}}
\newcommand{\qqbar}{q{\overline q}}
\newcommand{\ppbar}{p{\overline p}}
\newcommand{\gev}{\ts \rm GeV \ts}
\newcommand{\pico}{\ts \rm pb \ts}
\newcommand{\glu}{\cal G}
\newcommand{\mrhot}{m_{\rhoc}}
\newcommand{\meta}{m_{\etoc}}
\newcommand{\aqc}{\alpha_s}
\newcommand{\FQ}{F_Q}
\newcommand{\mtt}{M_{\ttbar}}
\newcommand{\shat}{\hat s}
\begin{document}
\begin{titlepage}
\begin{center}
\Large{{\bf  A New Physics Source of Hard Gluons \\in Top Quark
Production}}\footnote{Talk given at the 17th annual MRST meeting,
Rochester, May 8-9, 1995 --- based on work done with B.~Holdom. For a
more complete version of the results described here see \cite{paper}.}
\end{center}
\vspace{5mm}
\begin{center}
M.~V.~Ramana\footnote{e-mail address:
ramana@medb.physics.utoronto.ca}\\ {\normalsize\it Department of
Physics}\\ {\normalsize\it University of Toronto}\\ {\normalsize\it
Toronto, Ontario,}\\ {\normalsize CANADA, M5S 1A7}
\end{center}
\vspace{2cm}
\thispagestyle{empty}
\begin{abstract}
We consider the contribution of new strongly interacting sector, with a
characteristic scale of half a TeV, to top quark production at the
Tevatron.  The color-octet, isosinglet analog of the $\rho$ meson in
this theory is produced copiously in hadron colliders. If the mass of
this resonance is less than twice the mass of the lightest
pseudo-Goldstone bosons, then an important decay mode could be to the
color-octet analog of the $\eta$ and a gluon.  The subsequent decay of
this $\eta$ into $\ttbar$ gives rise to top quark events with a hard
gluon.
\end{abstract}
\end{titlepage}
\newpage

\renewcommand{\thefootnote}{\arabic{footnote}}
\setcounter{footnote}{0}
\setcounter{page}{2}

Recently the CDF and D0 collaborations have presented evidence confirming
the production of top quarks at the Tevatron Collider.  CDF \cite{cdfnew}
finds a top quark mass of $m_t = 176 \pm 8 \pm {10} \gev$ and the cross
section $\sigma(\ppbar \to \ttbar) = 6.8^{+3.6}_{-2.4} \pico$. D0 \cite{d0}
finds a mass of $m_t = 199^{+19}_{-21}\pm 22 \gev$ and a cross section $6.4
\pm 2.2 \pico$. For $m_t=176 \gev$ the QCD prediction for the cross section
is $\sigma(\ppbar \to \ttbar) = 4.79^{+0.67}_{-0.41} \pico$
\cite{laenen}.  While the CDF value for the cross section is not
inconsistent with the QCD prediction, it still leaves some room for
contributions to $\ttbar$ production from other sources. Thus one
would like to explore signals from new physics which may show up in
top quark studies.

In this talk we will consider the contribution of a color-octet,
isosinglet vector resonance, the $\rhoc$, to $\ttbar$ production. Such
a resonance is expected in any strongly interacting theory which has
at least one colored, electroweak doublet of fermions. For the most
part we consider $\rhoc$ masses in the 400-600 GeV range, although we
shall see that higher values may also be of interest.  Examples of
theories with masses in this range, which is somewhat below the usual
TeV electroweak symmetry breaking scale, are discussed in
\cite{multi},\cite{first}, and \cite{meta}.

As discussed in \cite{bob}, the $\rhoc$ has the odd-parity decay modes
$\rhoc \to \etoc \glu$ and $\rhoc \to \etas \glu$.  If the color-octet or
the color-singlet analogs of the QCD $\eta$ meson ($\etoc$ or $\etas$) are
above the $\ttbar$ threshold, then they decay predominantly into $\ttbar$
\cite{drk}. A signature to be explored in this talk is provided by the hard
gluon produced in addition to the $\ttbar$ pair.

The color-octet $\etoc$ itself can be strongly produced in hadronic
collisions and its contribution to $\ttbar$ production has been
considered by several authors \cite{drk} -- \cite{ken}.  However this
is mainly produced from initial states involving two gluons. At the
Tevatron, since we are dealing with processes where the initial
partons have a large fraction of the proton's momentum, the gluon
distribution functions are much smaller than the quark distribution
functions. This leads us to consider the $\rhoc$ which is produced
from $\qqbar$ initial states.  Since the production cross section for
a color-octet $\rhoc$ is quite large \cite{ehlq,first}, we expect that
the process $\qqbar \to \rhoc \to \etoc {\glu} \to \ttbar {\glu}$
could contribute substantially to the top quark production rate at the
Tevatron, depending on the masses of the $\rhoc$ and the $\etoc$.

There could also be a contribution to $\ttbar$ production from $\rhoc
\to \etas \glu$.  The color factors are such that the $\etas$ contribution
to the cross section is two-fifths the contribution of an $\etoc$ with
the same mass.  On the other hand for the mass ranges of the $\etoc$
that we will be exploring in this talk, the mass of the $\etas$ may
well be below the $\ttbar$ threshold. To be definite we will
henceforth assume that only the $\etoc$ contributes to the process of
interest.

The CDF results on the top quark mass and production cross section are
primarily based on detecting the $\ttbar$ pair in its leptons + jets
decay mode. This would nominally produce events with four jets.
However many of the events observed have five or more jets, as could
be expected from QCD gluon radiation. For such events the procedure
followed \cite{cdfnew} is to assume that the four highest $E_T$ jets
arise from the decay of the $t\overline t$ system. Thus even events
which have extra gluons are included in the CDF analysis, and hence
our mechanism for the production of $\ttbar \glu$ would contribute to
their measurement of the $\ttbar$ cross section.

\begin{table}
\centering
\begin{tabular}{|c|c|p{0.38in}|p{0.38in}|p{0.25in}|c|c|c|}\hline
\  $\mrhot$ \ &\ $\meta$ \  &\multicolumn{3}{c|}
{$\sigma({\ppbar \to \rhoc \to \ttbar + \glu})$}&\multicolumn{3}{c|}
{$\sigma({\ppbar \to \etoc \to \ttbar})$ }\\ [1mm] \hline
475 & 400 & 2.06 & 2.74 &2.99& 0.72 & 1.61 & 1.35 \\ [2mm]
475 & 450 & 0.71 & 0.72 &0.60& 0.42 & 0.83 & 0.40 \\ [2mm]
525 & 450 & 1.35 & 1.64 &1.69& 0.35 & 0.69 & 0.33 \\ [2mm]
750 & 400 & 0.25 & 0.41 &0.25& 0.35 & 0.72 & 0.34 \\ [2mm]
\hline
\end{tabular}
\label{mej1}

\caption{Cross sections for the production of $\ttbar + \glu$
at the Tevatron. For each set of masses, the first column under each
cross section corresponds to \{$N=2,k=1.0$\}, the second to
\{$N=4,k=1.0$\} and the third to \{$N=2,k=0.1$\}. All masses are in GeV
and all cross sections are in pb. No QCD corrections are included.}

\end{table}

Before presenting the cross sections for $\ttbar$ production, we first
show the partial widths for the various decay modes that are relevant
to our calculations.  The partial widths for the decay
modes\footnote{We are assuming that the $\rhoc$ is below the two
pseudo-Goldstone boson threshold.} of the $\rhoc$ can be calculated
using the ideas outlined in \cite{vmd,bky,mike} to be:
\beq
\Gamma(\rhoc \to \qqbar) + \Gamma(\rhoc \to {\glu \glu}) =
{5 \over {6}} \ts {\aqc^2(\mrhot) \over \alpha_{\rhoc}}\ts \mrhot +
{1 \over {2}} \ts {\aqc^2(\mrhot) \over \alpha_{\rhoc}}\ts \mrhot \ts,
\eeq
and
\beq
\Gamma(\rhoc \to \etoc {\glu}) = {5 \over {128}}{1 \over{\pi^3}}
{\left(N \over {3}\right)^2} {\alpha_{\rhoc} \aqc(\mrhot) \over
{\FQ^2}} {\left({{\mrhot^2 - \meta^2} \over \mrhot}\right)^3}
\ts
\eeq
Following \cite{ehlq}, we have assumed the following scaling relations
for $\alpha_{\rhoc}$, the $\rhoc$ decay constant and $\FQ$, the analog
of the pion decay constant
\bea
\alpha_{\rhoc} &=& {3 \over {N}} \ts 2.97 \nonumber \\
{\FQ \over \mrhot} &=& {f_{\pi} \over {m_{\rho}}} \ts \sqrt{{{N} \over
3}} \label{scaling}
\eea
where 2.97 is the QCD $\rho$ meson decay constant.  $N$ is the number of
``colors'' in this theory; if this were to be a technicolor theory then $N$
would be the number of technicolors.

The decay widths of the $\etoc$ are given by \cite{drk}
\beq
\Gamma (\etoc \to \ttbar) + \Gamma (\etoc \to {\glu \glu}) =
 k_t {{m_t^2 \meta} \over {16 \pi \FQ^2}} {\sqrt{\left(1-{{4 m_t^2}
\over {\meta^2}}\right)}} + {{5 N^2 \aqc \meta^3}
\over {384 \pi^3 \FQ^2}} \ts.
\eeq
The constant of proportionality for the $\ttbar$ decay, $k_t$, is
determined by some underlying theory and is expected to be of order
one. In our calculations, $k_t$ is varied between 0.1 and 1.  The
decay $\etoc \to \ttbar$ dominates over the decay into $\glu \glu$
except when $\meta$ is very close to the $\ttbar$ threshold or when
$k_t$ is very small.

\begin{figure}
\input{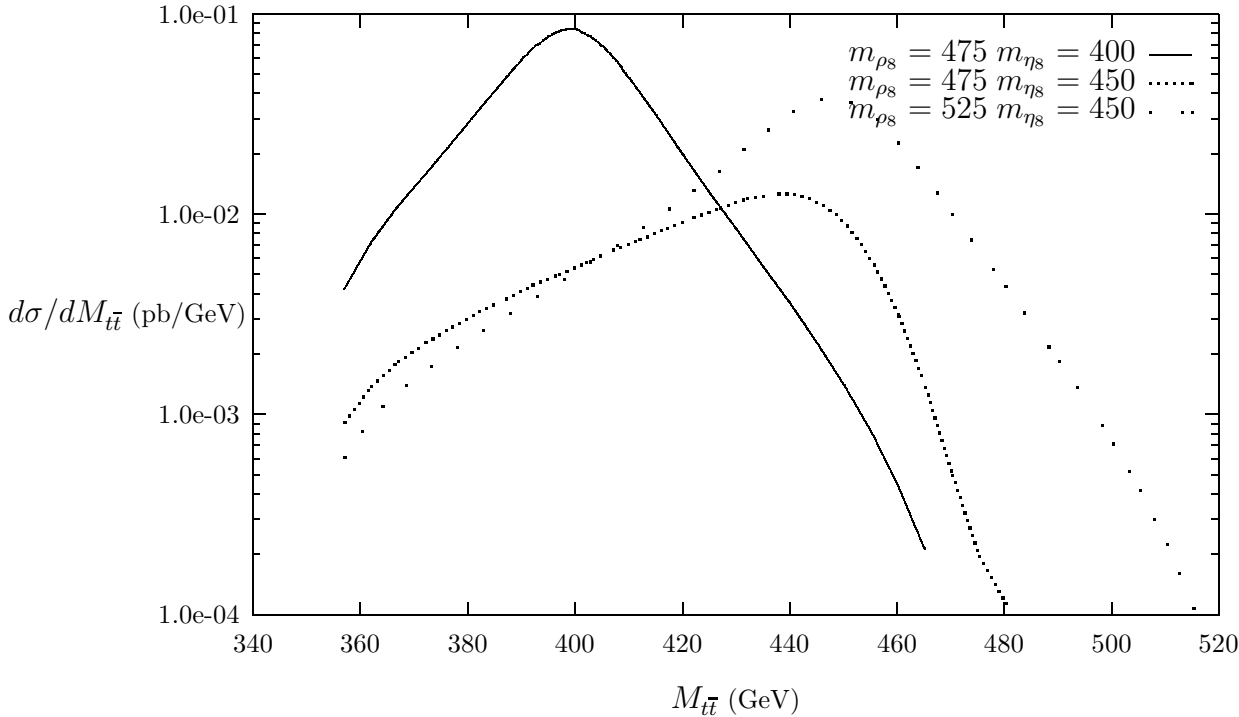}
\vspace{ 0.1 in}
\caption{The differential cross section for total $\ttbar + \glu$
production with $\mtt$ being the invariant mass of $\ttbar$.  No
rapidity cuts are imposed. $N = 2$ and $k_t = 1$.}
\label{dcsy92}
\end{figure}

The resulting partonic level differential cross section for the process
$\qqbar \to \ttbar {\glu}$ is
\bea
{{d{\hat \sigma}} \over {dz d\mtt}} &=& {5 \over {384 \pi^4}} \ts
{\left({N \over 3}\right)^2} \ts {{\aqc^3 k_t m_t^2} \over \FQ^4} \ts
{\mrhot^4 \over {((\shat-\mrhot^2)^2 + \shat \Gamma^2_{\rhoc})}}
\nonumber \\
& &\left(1 - {\mtt^2 \over \shat}\right)^2
\sqrt{\left(1-{{4 m_t^2} \over {\mtt^2}}\right)} \ts
{{(\mtt^2 - 2m_t^2) \mtt } \over {((\mtt^2 - \meta^2)^2 + \mtt^2
\Gamma^2_{\etoc})}} \ts
(1+ z^2) \ts.
\eea
where $\mtt$ is the invariant mass of the $\ttbar$ pair, $\shat$, the
partonic center of mass energy and $z = {\rm cos}(\theta)$ with
$\theta$ being the angle between the outgoing gluon and the incoming
quark.  In view of the complexity of the top quark analysis followed
by the CDF and D0 collaborations \cite{cdfnew,d0,cdf}, no rapidity
cuts are imposed on the top quarks. In order to be conservative, we
also do not include any QCD corrections although they are expected to
be significant, as in the case of $\ttbar$ production from QCD
\cite{laenen}.  These corrections could easily increase the production
rate by 50\% or more.

\begin{figure}
\input{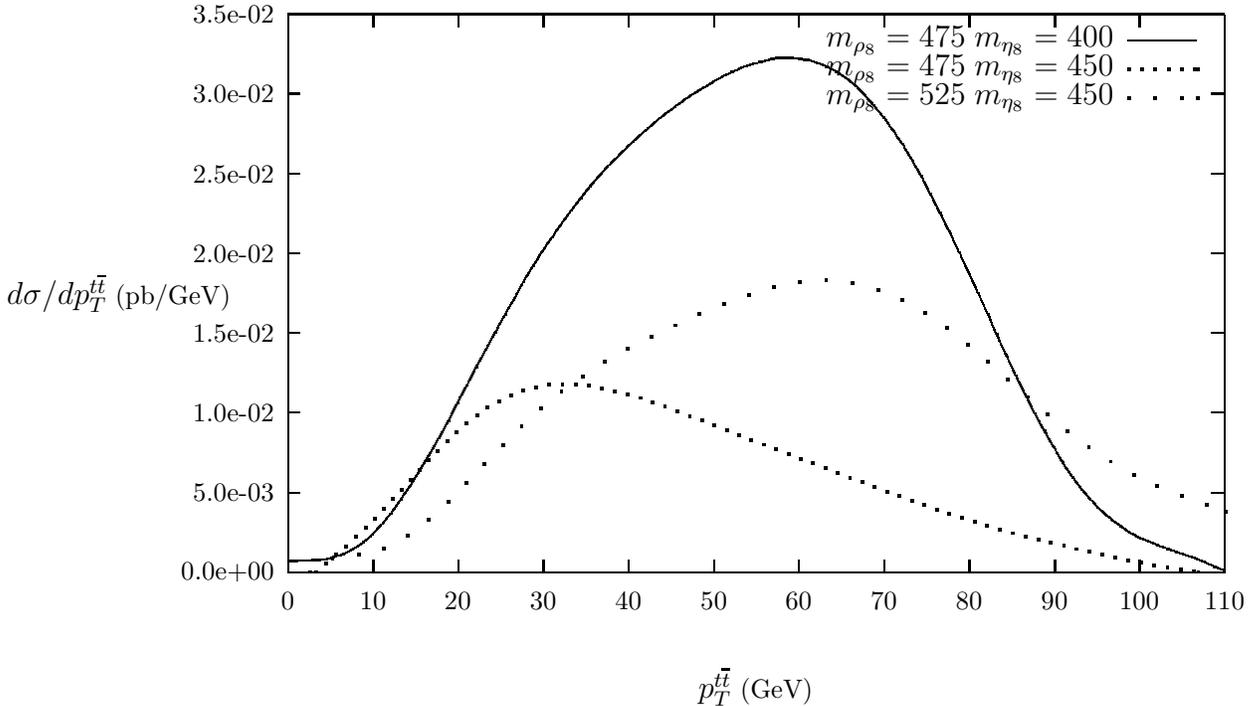}
\vspace{ 0.1 in}
\caption{The $p_T^{\ttbar}$ distribution in $\ttbar + \glu$
production.  A rapidity cut of $y_c = 2.0$ is imposed on the final
state gluon. $N = 2$ and $k_t = 1$.}
\label{gluet}
\end{figure}

This parton level result can be folded in with the quark distribution
functions and integrated over $z ={\rm cos}(\theta)$ and $\mtt$ in
order to obtain the total production cross section for $\ttbar +
{\glu}$.  We have ignored the contribution from ${\glu \glu} \to
\ttbar + \glu$ since the gluon structure functions are small at the
Tevatron.  The results of this computation are presented in Table~1
which shows the cross sections for $\ttbar$ production for different
values of $\mrhot$, $\meta$, $N$ and $k_t$.  We have used the EHLQ set
II structure functions \cite{ehlq} with $Q^2 = \hat{s}$ and have
chosen values of $\mrhot$ consistent with the range excluded in
\cite{dijet}.

The total cross section for $\ttbar$ production corresponds to events
where the extra (hard) gluon may or may not be detected. For
comparison, along with the cross sections for $\ppbar \to \rhoc \to
\ttbar + \glu$, we also show the contribution from $\ppbar \to \etoc
\to \ttbar$ using the formulae in \cite{ken} for the parameters we
have chosen.  However, unlike the authors of \cite{ken}, we do not
incorporate any estimates of QCD corrections.

The largest $\rhoc$ mass displayed in Table~1 is 750 GeV; this mass is
beginning to approach a value more typical of standard technicolor
theories. In this case, the contribution to top quark production from
our mechanism is fairly small (about 5\% or less of the estimated
cross section). However these events would have a unique signature ---
a $\ttbar$ pair with large $\peetee$ recoiling against a high energy
gluon.

For the lower $\mrhot$ values we consider in more detail methods to
distinguish our contribution to top quark production from the standard
model contribution. In order to do this we focus on three different
distributions --- the invariant mass $\mtt$ and the transverse
momentum $p_T^{\ttbar}$ of the $\ttbar$ pair, and the total invariant
mass $M_{inv}$ of $\ttbar +\glu$. These distributions are plotted
Figures (1-3) for the parameter values : $\{\mrhot = 475 \gev$, $\meta
= 400 \gev\}$, $\{\mrhot = 475 \gev$, $\meta = 450 \gev\}$ and
$\{\mrhot = 525 \gev$, $\meta = 450 \gev\}$.

The differential cross section ${{d{\hat \sigma}}/{d\mtt}}$ is
displayed in Fig.~\ref{dcsy92}. Since we are interested in estimating
the total top quark production, we have not imposed any rapidity cuts
on the gluon.  As expected, there is a peak in the $\mtt$ distribution
due to the intermediate resonance, the $\etoc$. However for heavier
$\etoc$, the $\mtt$ distribution is quite broad and the peak may not
show up very cleanly.

\begin{figure}
\input{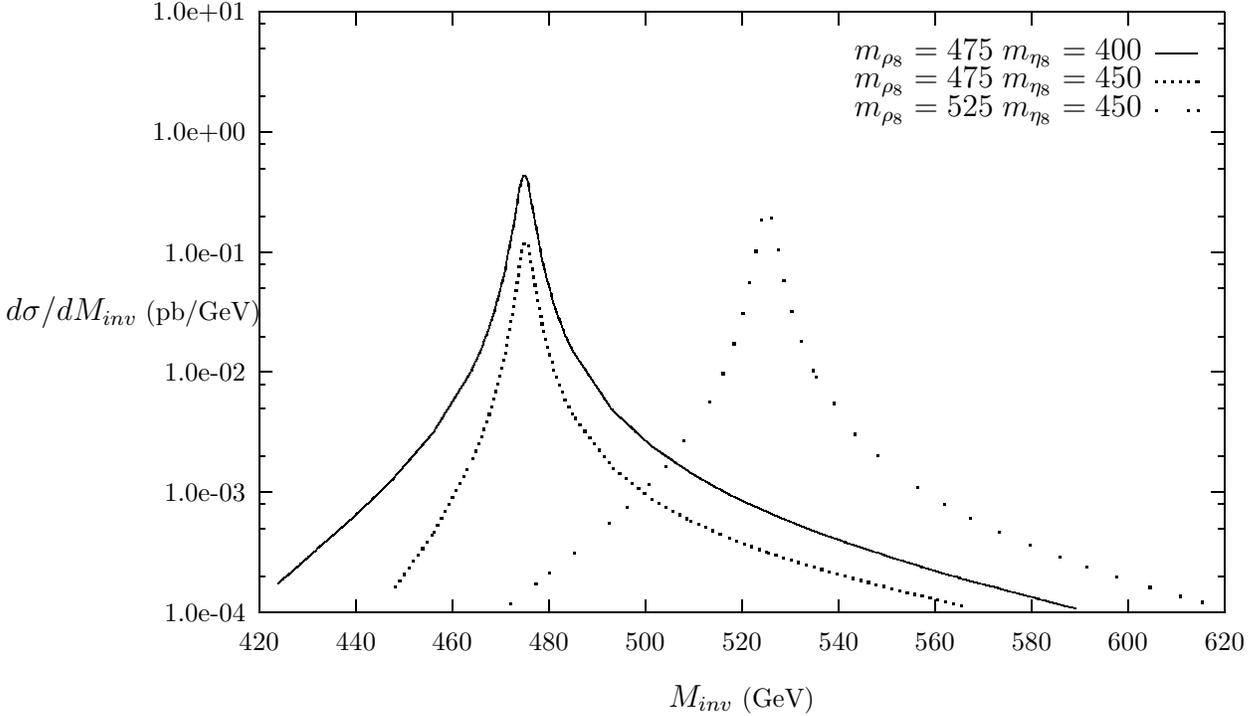}
\vspace{ 0.1 in}
\caption{The differential cross section for $\ttbar + \glu$
production with $M_{inv}$ being the invariant mass of $\ttbar + \glu$.
A rapidity cut of $y_c = 2.0$ is imposed on the final state gluon.  $N
= 2$ and $k_t = 1$.}
\label{dc3}
\end{figure}

Another variable of interest for new physics is the transverse
momentum of the $\ttbar$ pair $p_T^{\ttbar}$.  We show the
distribution from our mechanism in Fig.~\ref{gluet}. In calculating
this distribution (and the one in Fig.~\ref{dc3}) we impose a rapidity
cut of $y_c = 2.0$, so that the gluon be within the ambit of the
detector.  This is the value used by the CDF collaboration \cite{cdf};
we found that this cut reduced the total event rates by about ten to
fifteen percent. The standard model $p_T^{\ttbar}$ distribution has
been obtained \cite{nason} using both next-to-leading order QCD
calculations and HERWIG simulations. Similarly the $p_T$ distribution
for gluons radiated in $\ttbar$ production has been calculated
\cite{orr}. The distribution peaks for $p_T^{\ttbar}$ around 5 GeV and
then falls rapidly for increasing $p_T^{\ttbar}$. A significant excess
of high $p_T$ $\ttbar$ pairs would signal a nonstandard production
mechanism.

The $p_T^{\ttbar}$ distribution is closely related to the distribution
of the quantity referred to as $X$ in \cite{cdf}.  This is the total
{\it observed} transverse energy remaining in the event after
subtracting vectorially the transverse energy of the four jets and the
charged lepton.  The $X$ values for the seven events used for mass
fitting in \cite{cdf} are \{7.1~GeV, 7.7~GeV, 14.8~GeV, 17.0~GeV,
20.0~GeV, 26.3~GeV and 35.6~GeV\}.  A larger data set is clearly
required before any serious analysis can be made.

We notice that there is a significant bias in the extraction of $X$
which distorts its interpretation as a measure of $p_T^{\ttbar}$. The
method \cite{cdfnew} used in deciding which of the jets in an event
come from top decays is to assign the four highest energy jets to the
decay of the $\ttbar$ system.  While this is suitable for standard
model $\ttbar$ production, it may not be appropriate if one wishes to
allow for new physics. In particular, the hard gluon produced in our
mechanism can be mistakenly assigned as one of the jets from the
$\ttbar$ system.  This in turn could distort the extraction of the top
mass.  It also implies that the observed $X$ distribution is likely to
be softer than the true $p_T^{\ttbar}$ distribution.

The third variable we consider is the total invariant mass of the
event, which in our case is the invariant mass of the $\ttbar +\glu$.
Since the $\rhoc$ is relatively narrow this distribution is sharply
peaked, as can be seen in Fig.~\ref{dc3}. This variable is also less
prone to experimental ambiguities involved in assigning jets to
underlying partons.

In conclusion, we have suggested a mechanism for the production of top
quarks at the Tevatron. The $\rhoc$ and $\etoc$ resonances which take
part in this resonant enhancement of $\ttbar$ events are generic of
any strongly interacting theory with at least one colored, electroweak
doublet of fermions. If these resonances have masses in the range of
400-550~GeV range, they can contribute significantly to $\ttbar$
production at the Tevatron.  We have suggested ways of distinguishing
this mechanism from standard model production.  We finally note that
if the color-singlet $\etas$ was also above the top-pair threshold
then we would have contributions from both the $\etoc$ and the
$\etas$. There would be two peaks in the $M_{\ttbar}$ distribution,
with their relative contributions being fixed by the masses of the
resonances.

{\bf Acknowledgements}

We would like to thank P.~Sinervo and G.~Triantaphyllou for useful
discussions.  This research was supported in part by the Natural
Sciences and Engineering Research Council of Canada. We would like to
thank the organizing committee of the MRST conference for the
opportunity to present these results.


\begin{thebibliography}{99}

\bibitem{paper} B.~Holdom and M.~V.~Ramana, University of Toronto Preprint,
UTPT-95-07,\\ hep-ph/9504403, to appear in \pl.

\bibitem{cdfnew} The CDF Collaboration, ``Observation of Top Quark
Production in $\ppbar$ Collisions'', FERMILAB-PUB-95/022,
hep-ex/9503002.

\bibitem{d0} The D0 Collaboration, ``Observation of the Top Quark'',
FERMILAB-PUB-95-028-E, hep-ex/9503003

\bibitem{laenen} E.~Laenen, J.~Smith and W.~L.~van Neerven, \pl {\bf 321}
(1994) 254.

\bibitem{multi} K. Lane and E. Eichten, \pl~{\bf 222} (1989) 274.

\bibitem{first} K. Lane and M. V. Ramana, \pr~{\bf D44} (1991) 2678.

\bibitem{meta} B.~Holdom, \pl {\bf 314} (1993) 364;
 \pl {\bf 336} (1994) 85\semi
S.~F.~King, \pl {\bf 314} (1993) 364.

\bibitem{bob} B.~Holdom, \pl {\bf 344} (1995) 355.

\bibitem{drk} S.~Dimopoulos, S.~Raby and G.~L.~Kane, \np {\bf 182}
(1981) 77.

\bibitem{girardi} G.~Girardi, P.~Mery and P.~Sorba,
\np {\bf 195} (1982) 410.

\bibitem{ehlq} E. Eichten, I. Hinchliffe, K. Lane and C. Quigg,
\rmp~{\bf 56} (1984) 579.

\bibitem{baltay} C.~Baltay and G. ~Segre, in {\it Proceedings
of the 1984 Summer Study on the Design and Utilization of the
Superconducting Supercollider, Snowmass, 1984}, edited by R.~Donaldson
and J.~Morfin (Division of Particles and Fields of the American
Physical Society, New York, 1985) 299.

\bibitem{kuo} W.~C.~Kuo, B.~L.~Young and D.~W.~McKay, \pr {\bf D36}
(1987) 2729.

\bibitem{george} T.~Appelquist and G.~Triantaphyllou, \prl {\bf 69}
(1992) 2750.

\bibitem{ken} E.~Eichten and K.~Lane, \pl {\bf 327} (1994) 129.

\bibitem{vmd} J.~J.~Sakurai, {\it Currents and Mesons} (1969)
University of Chicago Press.

\bibitem{bky} M.~Bando, T.~Kugo and K.~Yamawaki, \prp {\bf 164} (1988) 217.

\bibitem{mike} A.~Falk and M.~Luke, \np {\bf 337} (1990) 49.

\bibitem{cdf} The CDF collaboration, \pr~{\bf D50} (1994) 2966;
 \prl {\bf 73} (1994) 225.

\bibitem{dijet} The CDF Collaboration, \prl {\bf 74} (1995) 3538.

\bibitem{nason} S.~Frixione, M.~L.~Mangano, P.~Nason and G.~Ridolfi,
``Top Quark Distributions in Hadronic Collisions'',
CERN-TH/95-52, GeF-TH-3/1995, hep-ph/9503213.

\bibitem{orr} L.~H.~Orr, T.~Stezler and W.~J.~Stirling,
``Gluon Radiation in T Anti-T production at the Tevatron P Anti-P
collider'', DTP-94-112, hep-ph/9412294.

\end{thebibliography}
\end{document}